\documentstyle[12pt]{article}

\catcode`\@=11
\long\def\@makefntext#1{
\protect\noindent \hbox to 3.2pt {\hskip-.9pt
$^{{\ninerm\@thefnmark}}$\hfil}#1\hfill}                %CAN BE USED

 \def\@makefnmark{\hbox to 0pt{$^{\@thefnmark}$\hss}}  %ORIGINAL

\def\ps@myheadings{\let\@mkboth\@gobbletwo
\def\@oddhead{\hbox{}
\rightmark\hfil\ninerm\thepage}
\def\@oddfoot{}\def\@evenhead{\ninerm\thepage\hfil
\leftmark\hbox{}}\def\@evenfoot{}
\def\sectionmark##1{}\def\subsectionmark##1{}}

\newcounter{sectionc}\newcounter{subsectionc}\newcounter{subsubsectionc}
\renewcommand{\section}[1] {\vspace{0.6cm}\addtocounter{sectionc}{1}
\setcounter{subsectionc}{0}\setcounter{subsubsectionc}{0}\noindent
        {\bf\thesectionc. #1}\par\vspace{0.4cm}}
\renewcommand{\subsection}[1] {\vspace{0.6cm}\addtocounter{subsectionc}{1}
        \setcounter{subsubsectionc}{0}\noindent
        {\it\thesectionc.\thesubsectionc. #1}\par\vspace{0.4cm}}
\renewcommand{\subsubsection}[1]
{\vspace{0.6cm}\addtocounter{subsubsectionc}{1}
        \noindent {\rm\thesectionc.\thesubsectionc.\thesubsubsectionc.
        #1}\par\vspace{0.4cm}}

\newcounter{appendixc}
\newcounter{subappendixc}[appendixc]
\newcounter{subsubappendixc}[subappendixc]

\renewcommand{\appendix}[1] {\vspace{0.6cm}
        \refstepcounter{appendixc}
        \setcounter{figure}{0}
        \setcounter{table}{0}
        \setcounter{equation}{0}
        \renewcommand{\thefigure}{\Alph{appendixc}.\arabic{figure}}
        \renewcommand{\thetable}{\Alph{appendixc}.\arabic{table}}
        \renewcommand{\theappendixc}{\Alph{appendixc}}
        \renewcommand{\theequation}{\Alph{appendixc}.\arabic{equation}}
        \noindent{\bf Appendix \theappendixc #1}\par\vspace{0.4cm}}

\def\abstracts#1{{
        \centering{\begin{minipage}{30pc}\tenrm\baselineskip=12pt\noindent
        \centerline{\tenrm ABSTRACT}\vspace{0.3cm}
        \parindent=0pt #1
        \end{minipage}}\par}}

\renewenvironment{thebibliography}[1]
        {\begin{list}{\arabic{enumi}.}
        {\usecounter{enumi}\setlength{\parsep}{0pt}
\setlength{\leftmargin 1.25cm}{\rightmargin 0pt}
         \setlength{\itemsep}{0pt} \settowidth
        {\labelwidth}{#1.}\sloppy}}{\end{list}}

\newcounter{itemlistc}
\newcounter{romanlistc}
\newcounter{alphlistc}
\newcounter{arabiclistc}

\newcommand{\fcaption}[1]{
        \refstepcounter{figure}
        \setbox\@tempboxa = \hbox{\tenrm Fig.~\thefigure. #1}
        \ifdim \wd\@tempboxa > 6in
           {\begin{center}
        \parbox{6in}{\tenrm\baselineskip=12pt Fig.~\thefigure. #1}
            \end{center}}
        \else
             {\begin{center}
             {\tenrm Fig.~\thefigure. #1}
              \end{center}}
        \fi}

\newcommand{\tcaption}[1]{
        \refstepcounter{table}
        \setbox\@tempboxa = \hbox{\tenrm Table~\thetable. #1}
        \ifdim \wd\@tempboxa > 6in
           {\begin{center}
        \parbox{6in}{\tenrm\baselineskip=12pt Table~\thetable. #1}
            \end{center}}
        \else
             {\begin{center}
             {\tenrm Table~\thetable. #1}
              \end{center}}
        \fi}

\def\@citex[#1]#2{\if@filesw\immediate\write\@auxout
        {\string\citation{#2}}\fi
\def\@citea{}\@cite{\@for\@citeb:=#2\do
        {\@citea\def\@citea{,}\@ifundefined
        {b@\@citeb}{{\bf ?}\@warning
        {Citation `\@citeb' on page \thepage \space undefined}}
        {\csname b@\@citeb\endcsname}}}{#1}}

\newif\if@cghi
\def\cite{\@cghitrue\@ifnextchar [{\@tempswatrue
        \@citex}{\@tempswafalse\@citex[]}}
\def\citelow{\@cghifalse\@ifnextchar [{\@tempswatrue
        \@citex}{\@tempswafalse\@citex[]}}
\def\@cite#1#2{{$\null^{#1}$\if@tempswa\typeout
        {IJCGA warning: optional citation argument
        ignored: `#2'} \fi}}

\def\fnt#1#2{\footnotetext{\kern-.3em
        {$^{\mbox{\sevenrm #1}}$}{#2}}}

 1
 1
 1

\font\tenbf=cmbx10
\font\tenrm=cmr10
\font\tenit=cmti10

\font\ninerm=cmr9

\newcommand{\be}{\begin{equation}}
\newcommand{\ee}{\end{equation}}

\newcommand{\eps}{\epsilon}

\newcommand{\R}{I\!\! R}

\begin{document}

\baselineskip=22pt
\centerline{\tenbf CONFINEMENT IN 3D GLUODYNAMICS }
\centerline{\tenbf AS A 2D CRITICAL PHENOMENON}
\baselineskip=16pt
\vspace{0.5cm}
\centerline{\tenrm
A. FERRANDO$ ^a$, A. JARAMILLO$ ^b$  and S. V. SHABANOV$ ^b$
\footnote{on leave from {\em Laboratory of Theoretical Physics,
JINR, Dubna, Russia.}}}
\baselineskip=13pt
\centerline{\tenit $ ^a$Center for Theoretical Physics,
Laboratory for Nuclear Science and Department of Physics}
\centerline{\tenit Massachusetts Institute
of Technology, Cambridge, Massachusetts 02139}
\baselineskip=13pt
\centerline{\tenit $ ^b$ Departament de F\'{\i}sica Te\`{o}rica and I.F.I.C.
Centre Mixt Universitat de Val\`{e}ncia -- C.S.I.C.}
\baselineskip=12pt
\centerline{\tenit E-46100 Burjassot (Val\`{e}ncia), Spain.}
\vspace{0.3cm}
\abstracts{
Gluodynamics in 3D spacetime with one spatial direction
compactified into a circle of length $L$ is studied.
The confinement order parameters, such
as the Polyakov loops, are analyzed in both the limits
$L\rightarrow 0$ and $L\rightarrow \infty$.
In the latter limit the behavior of the confinement
order parameters is shown to be described by a 2D
non-linear $\sigma$-model on the compact coset space
$G/ad\ G$, where $G$ is the gauge group and
$ad\ G$ its adjoint action on $G$. Topological
vortex-like excitations of the compact field variable cause a
Kosterlitz-Thouless phase transition which is
argued to be associated with the confinement phase transition
in the 3D gluodynamics.}

\vspace{0.2cm}
\rm\baselineskip=14pt

\noindent
{\bf 1. Introduction.}
In recent years there exists an increasing
interest in relating some $QCD$ non-perturbative
phenomena to 2D field theories properties\cite{Adjointmatter}.
A way to describe
a relation between 3D gluodynamics and its effective 2D field
theories
is to compactify one of the two {\em spatial} directions into
a circle of length $L$.
The length $L$ plays the role of
an interpolating parameter between the 2D and 3D gluodynamics
($L \rightarrow
0$ and $L \rightarrow \infty$, respectively).
When $L$ goes to zero, gauge fields homogeneous in the
compactified coordinate will dominate because the non-homogeneous
components of the gauge fields become massive (the mass is of order
$1/L$). The effective theory in this limit is $QCD_2$ with the
adjoint scalar matter\cite{dh82,Adjointmatter,fj95}.

Our interest lies on both the study
of the relevant order parameters describing 3D confinement and
constructing 2D models which can explain their critical behaviors.
The sought effective 2D theory should therefore be valid beyond
the small $L$ regime. The main aim of the paper is to describe such
a theory.

\vspace{0.2cm}
\noindent
{\bf 2. Characterization of $QCD_3$ confinement in 2D.}
One can define the transverse
(along the compactified direction) and the longitudinal
(along the unbounded directions) Polyakov loops as
the {\em v.e.v.}'s of the 2D $SU(N)$ field operators:
\begin{eqnarray}
P \equiv  \lim_{\beta \rightarrow \infty} \langle
\mbox{Tr} g(0) \rangle_\beta
&, &
 g(x_1,x_2) = \exp[i g_3\textstyle\int_0^\beta
 \!\! dx_0  A_0 (x_0,x_1,x_2)]
     \in SU(N)\ ;
                                        \nonumber \\
\bar{P}(L) \equiv \langle \mbox{Tr} \bar{g} (0)\rangle_L &, &
         \bar{g} (x_0,x_1) = \exp[i g_3 \sqrt{L} \phi (x_0,x_1)]
    \in SU(N)\ ; \nonumber \\
        & &
\phi (x_0,x_1) \equiv L^{-1/2}
\textstyle\int_0^L \!\! dx_2  A_2 (x_0,x_1,x_2)\ .
\label{Polyakov}
\end{eqnarray}
Since $g$ and $\bar{g}$ are 2D operators,
the behavior of the Polyakov loops can
be inferred from pure 2D considerations.

In the small $L$ limit exact calculations of these loops are possible.
This regime is characterized by $(P\!=\!0, \bar{P} \!\neq\! 0)$.
The other limit
is expected to be given by $(P\!=\!0, \bar{P} \!= \!0)$.
Therefore, although both regimes
are in a confining phase ($P\!=\!0$) they are not in the same {\em chiral}
phase yet.
The non-zero value of the transverse Polyakov loop
($\bar{P} \!\neq\! 0$) indicates that the small $L$
regime is in a chirality broken phase. When $L$ grows,
the timelike string tension (the one associated with $P$)
stabilizes with $L$ at the same critical length as $\bar{P}$ vanishes
as lattice simulations show \cite{Lattice}.

The physical interpretation of this result is very appealing.
For small values of
$L$, there is no difference in the confinement picture occurring
in $QED$ and $QCD$ because color flux lines are
forced to lie on the
artificial strip of a small width $L$
yielding a linearly rising potential between static charges.
However the distribution of flux lines on the
strip behaves very differently in both cases. In the abelian case flux
lines spread out occupying
the whole width independently of the strip size.
In the non-abelian case the transverse gluon interactions
increase the string tension
value by keeping the flux lines more and more packed.
When the critical length is reached the flux lines remains together
independently of how much the strip width is enlarged.
The fact that this phenomenon
occurs at the same time as
$\bar{P}$ vanishes
tells us that we can use the transverse Polyakov
loop to characterize the generation of the flux tube.

{}From a purely 2D point of view,
the problem of the flux tube generation boils down
in the understanding of the 2D field $\phi$ dynamics.
Our strategy is to construct a consistent 2D model incorporating
two important properties. Firstly, it must reduce to the known
small width regime result which can be calculated exactly from $QCD_3$.
Secondly, it has to exhibit a non-trivial 2D phase
transition responsible for the vanishing
of the {\em v.e.v.} of the 2D operator $\mbox{Tr} \bar{g}$.

\vspace{0.2cm}
\noindent
{\bf 3. 2D effective theory.}
Our starting point is to perform
the Fourier decomposition of the gauge field over
the compactified coordinate $x_2$ and thereby to obtain
an equivalent 2D theory. We choose the gauge
$\partial_2 A_2 \!=\! 0$ or $A_2(x,x_2)\! =\!\phi(x)/\sqrt{L}$,
in order for the theory to be invariant
under gauge transformations depending
only on the longitudinal coordinates $(x_0,x_1)$.
The gauge fixed action will be a function of the infinite set of
the longitudinal modes $(\phi, a_\alpha, \{V_\alpha^n\})$. The
non-zero mode fields $\{V_\alpha^n\}$ are massive, with bare masses
proportional to $1/L$.
The zero mode fields $(\phi, a_\alpha)$ are classically massless.

An effective action for the $\phi$ field, which allows
us to calculate $\langle \mbox{Tr} \bar{g} \rangle$, is obtained by
integrating out the non-zero modes.
No matter how complicated the interaction is, the
effective action has to possess two important properties. First of all,
it has to enjoy the longitudinal gauge invariance
for the 2D zero mode fields.
The field $a_\alpha$ plays the role of the 2D gluon field,
whereas $\phi$ is the adjoint scalar (matter) field.
The second condition is that the $\phi$ field is not an ordinary
Lie-algebra-valued field, rather it parametrizes
the {\em compact} gauge group manifold $G$.
Accordingly, the functional integration
over the field $\phi$ involves a nontrivial local
measure $d\phi M(\phi) = d\mu_H(\bar{g})$ with
$d\mu_H$ being the left- and right-invariant Haar measure for
$\bar{g}(x_0,x_1)\in G$.
The origin of this dynamical condition
on the field $\phi$ is easily understood if one calculates the
Faddeev-Popov determinant in the gauge $A_2=\phi/\sqrt{L}$,
which coincides with the Haar measure\cite{sh93}.
For instance for SU(2) it reads $M(\phi)=
\sin^2(\eps |\phi|)/|\phi|^2$, where
$|\phi|= \sqrt{{\rm Tr}\phi^2}$, $\eps=g_3 \sqrt{L} /\sqrt{2}$,
$\phi = \tau_a\phi_a/2$ and ${\rm Tr} \tau_a\tau_b =2\delta_{ab}$
for the Pauli matrices.

When integrating out the gluon field $a_\alpha$,
we fix the unitary gauge $\phi=\tau_3\varphi/2$. The Faddeev-Popov
determinant for this gauge is $\varphi^2$ so the final measure
for the field $\varphi$ assumes the form $\sin^2(\varphi \eps)$.
The effective 2D action for
the field $\varphi$ should respect the discrete symmetry
$T_n:\, \varphi \rightarrow \pm \varphi + 2 \pi n\eps^{-1} $,
(where $n$ is an integer) which is known as the
affine Weyl group $W_A$ and can be viewed as a set of
residual gauge transformations that are still allowed by the gauge
$A_2 = \tau_3\varphi/2$ \cite{sh93}, that is, the physical
values of $\varphi$ lie in the interval $(0, \pi\eps^{-1})$ between
two nearest zeros of the Faddeev-Popov determinant.

>From the mathematical point of view, the effective
2D theory should be a non-linear
$\sigma$-model on the coset space $G/ad\ G$, where $ad\ G$ is
the adjoint action of the gauge group on the group manifold $G$.
Note that the Lie-algebra-valued field $\phi$
parametrizes $G$ and realizes the adjoint representation of the
gauge group. Therefore its gauge non-equivalent configurations
form the coset space $G/ad\ G$. To parametrize them,
it is natural to choose a gauge
where $\phi$ belongs to the Cartan subalgebra $H$.
Then $G/ad\ G$ is isomorphic not to
the whole Cartan subalgebra $H$, but to a compact domain in
it $H/W_A$ known as the Weyl cell\cite{sh93}.
In the $SU(2)$ case $H$ is isomorphic to a line $\R$ and, hence,
the Weyl cell is $(0,\pi\eps^{-1}) = \R /T_n$.

\vspace{0.15cm}
\noindent
{\bf 4. Flux tube formation as a Kosterlitz-Thouless transition.}
It is certainly not possible to find an explicit form of the effective
action
for the compact (cyclic) field variable $\varphi$.
However, one can model its critical behavior.
The field $\varphi$ is compact so it should have
topological vortex-like excitations that could cause the Kosterlitz-Thouless
phase transition\cite{kt73,zin}. To describe the dynamics of vortices,
the compact field $\varphi$ has to be mapped on a non-compact field
$\vartheta$ in the partition function path integral.
To regularize the theory
at short distances, we set the system on a 2D lattice.
The variable $\eps\varphi_x\in (-\pi,\pi)$
at the lattice site $x$ can be thought
as an angular variable determining a direction of spin attached
to $x$. That is, the effective theory describes a planar spin system.
Applying a conventional technique\cite{kadanoff} we get
\begin{eqnarray}
{\cal Z} &=& \textstyle{\int_{-\pi/\eps}^{\pi/\eps}
\prod_x} \left(\eps d\varphi_x \,
\sin^2(\varphi_x \eps)\right) \,
\exp\left\{-\textstyle{\sum_{\langle x,x'\rangle}} {\cal L}(\varphi_x,
\varphi_{x'})\right\}
\label{2} \\
&=&\textstyle{ \sum_{[m_x]} \int_{-\infty}^{\infty} \prod_x} (
d\vartheta_x)
 \, \exp\left\{-\textstyle{\sum_{\langle x,x'\rangle }} \left(
 \tilde{\cal L}_M(\vartheta_x,
\vartheta_{x'},\eps)+2\pi i m_x \vartheta_x \right)\right\}
\label{3}
\end{eqnarray}
where $m_x$ is an integer-valued field (vorticity
at a site $x$), ${\cal L}(\varphi_x,\varphi_{x'})$
is a lattice Lagrangian of the cyclic field $\varphi_x$
and $\langle x,x'\rangle $
specifies lattice sites involved in a pair-wise
interaction.\footnote{
One may also assume ${\cal L}$ to contain non-local
interactions, not just nearest-neighbor interactions.}
The transformation (\ref{3}) is the duality transformation
that relates the strong coupling regime $\eps\rightarrow \infty$
of the model (\ref{2}) to a weak coupling regime in  (\ref{3}).

In the Villain model\cite{zin,kadanoff}, $\tilde{\cal L}_M$ is
quadratic and the integral over the non-compact spin-wave field $\vartheta$
can be done so that the partition function assumes the form of that
for the $2D$ Coulomb gas, where  $m_x$ plays the role of the charge
distribution. The model (\ref{2}) exhibits the Kosterlitz-thouless phase
transition which, and it is our conjecture, can be associated
with the stabilization of the flux tube at a certain critical
width $L=L_c$.
Indeed, the small $L$ regime ($|\varphi_x|<\!\! < \eps^{-1} $) corresponds
to a low temperature regime ($\eps^2\rightarrow 0$) in (\ref{2}).
Below a certain critical temperature the spin system is in an
"ordered phase"; spin-wave excitations are dominant, while vortices
are bound in vortex-antivortex pairs. Above the critical
temperature these pairs are expected to dissolve, which
results in vanishing the {\em v.e.v.} $\langle Tr\,\bar g \rangle $.

 To obtain a soluble (or analytically tractable) model, one should
take into account two important features of our effective theory which
are not present in the Kosterlitz-Thouless model and have the {\it gauge}
origin. First, in addition to the local periodicity $\varphi_x\rightarrow
\varphi_x + 2\pi \eps^{-1} n_x$, the Lagrangian ${\cal L}$ should
respect a {\it local} $Z\!\!\!Z_2$ symmetry $\varphi_x\rightarrow
\pm\varphi_x$, the residual gauge transformation from the Weyl group
(not to be confused with the
transformations from the center $Z\!\!\!Z_2$
of $SU(2)$, the later ones correspond to $\varphi_x\rightarrow
\varphi_x +\pi$). Thus, our system is a planar spin system with an
"extra" $Z\!\!\!Z_2$-gauge group.
In particular, this imposes a condition on
the possible forms of $\tilde{\cal L}_M$: $\tilde{\cal L}_M(\vartheta_x,
\vartheta_{x'})=\tilde{\cal L}_M(-\vartheta_x,
\vartheta_{x'})=\tilde{\cal L}_M(\vartheta_x,
-\vartheta_{x'})$. The second feature
is the local measure $M=\prod_x \sin^2(\varphi_x \eps)$ (the
Faddeev-Popov determinant). One can consider it as a crystalline
magnetic field in the spin system: ${\cal L} \rightarrow
{\cal L}+ V_c,\ V_c = \gamma \ln M = -\gamma
\sum_{n=1}^\infty n^{-1}(\cos 2\varphi_x\eps)^n$. Its effect is in
a shift of the critical temperature\cite{kadanoff}.

\vspace{0.2cm}
\noindent
{\bf 5. Acknowledgments.}
A.J. would like to thank V. Branchina, M. Caselle, T. Heinzl and J. Polonyi 
for interesting discussions, and a conference
grant from the European Union as well as a support of
a doctoral fellowship from IVEI (Valencian Institution for Research and
Studies, Spain) are gratefully acknowledged.
A.F. has maintained illuminating
conversations with S. Chandrasekharan, A.V. Matytsin and
P. Zaugg. S.V.S. thanks the Department of Theoretical Physics
of University of Valencia for the warm hospitality.

\vspace{0.2cm}
\noindent
{\bf 6. References}


\begin{thebibliography}{9}

\bibitem{dh82}
E. D'Hoker, Nucl.Phys. {\bf B201} (1982) 401.
\bibitem{Adjointmatter}
 S. Dalley and I.R. Klebanov, Phys. Rev {\bf D47} (1993) 2517;\\
G. Bhanot, K. Demeterfi, and I.R. Klebanov, Phys.
Rev {\bf D48} (1993) 4980;\\
D. Kutasov, Nucl. Phys {\bf B414} (1994) 33; \\
F. Antonuccio and S. Dalley, hep-lat/9505009, hep-ph/9506456.
\bibitem{fj95}
A. Ferrando and A. Jaramillo, Nucl.Phys.{\bf 457B} (1995) 57;
Phys.Lett. {\bf 341B} (1995) 342.
\bibitem{Lattice}
M. Teper, Phys.Lett. {\bf B311} (1993) 223; \\
G.S. Bali, J. Fingberg, U.M. Heller, F. Karsch, K. Schilling,
Phys. Rev. Lett. {\bf 71} (1993) 3059;\\
M. Caselle, R. Fiore, F. Gliozzi, P. Guaita and S. Vinti,
Nucl. Phys. {\bf B422} (1994) 397.
\bibitem{sh93} S.V. Shabanov, Phys.Lett.{\bf 318B} (1993) 323;
Commun.Theor.Phys.{\bf 4} (1995) 1.
\bibitem{kt73}
J. M. Kosterlitz and D. J. Thouless, J.Phys. {\bf C6} (1973) 1181.
\bibitem{zin}
J. Zinn-Justin, {\it Quantum Field Theory and Critical Phenomena}
(Clarendon, Oxford, 1990) (see pp. 671-672).
\bibitem{kadanoff}
J.V. Jos\'e, L.P. Kadanoff, S. Kirkpatrick and D.R. Nelson,
Phys.Rev. {\bf B16} (1977) 1217.

\end{thebibliography}
\end{document}